\documentclass[epj,referee]{svjour}
\usepackage{graphicx}
\usepackage{amssymb}

\begin{document}

\titlerunning{Micro-traps from permanent magnetic \\
film with in-plane magnetization}
\authorrunning{I.Barb author et al.}

\title{Creating Ioffe-Pritchard micro-traps from permanent magnetic \\
film with in-plane magnetization}
\author{I. Barb \and R. Gerritsma \and Y.~T. Xing \and J.~B. Goedkoop \and R.~J.~C. Spreeuw}
\institute{Van der Waals - Zeeman Institute, University of
Amsterdam, Valckenierstraat 65, 1018 XE  Amsterdam, The
Netherlands
\\ \email{ibarb@science.uva.nl }}


\abstract{We present designs for Ioffe-Pritchard type magnetic
traps using planar patterns of hard magnetic material. Two samples
with different pattern designs were produced by spark erosion of
40 $\mu$m thick FePt foil. The pattern on the first sample yields
calculated axial and radial trap frequencies of 51~Hz and 6.8~kHz,
respectively. For the second sample the calculated frequencies are
34~Hz and 11~kHz. The structures were used successfully as a
magneto-optical trap for $^{87}$Rb and loaded as a magnetic trap.
A third design, based on lithographically patterned 250 nm thick
FePt film on a Si substrate, yields an array of 19 traps with
calculated axial and radial trap frequencies of 1.5 kHz and 110
kHz, respectively. }

\maketitle

\section{Introduction}

Planar microstructures have recently emerged as a very powerful
and attractive tool for handling neutral ultracold atoms. These
so-called "atom chips" have been used to construct miniature atom
optical elements including traps, waveguides, and beamsplitters
\cite{FolKruCas00,ReiHanHan99,FolKruHen02,Rei02,DekLeeLor00}. Atom
chips are now in development to miniaturize atom optics and to
make it more robust. They have great potential for application in
quantum information processing
\cite{TreHomRei04,Q:CalHinZol00,Q:JanVidCir03} and in atom
interferometry. Large magnetic field gradients and thus large
trapping forces can be achieved on-chip, removing the necessity of
large, high power external coils. Miniature current carrying wires
can produce tight magnetic trapping potentials that have
successfully been used to create Bose-Einstein condensates of
rubidium atoms \cite{HanHomHan01,OttForSch01}.

Most atom chips so far have relied on a pattern of
current-carrying wires to generate the required magnetic field
gradients. Here we investigate a promising alternative based on
patterned hard magnetic films. Waveguide designs using
out-of-plane magnetized film have been reported before using
TbFeCo \cite{SidMcLHan01} and CoCr films \cite{Dav99,SidMcLHan02}
and magneto-optically patterned Co/Pt thin films
\cite{EriRamHin04}. In plane magnetized videotape has been used
for waveguides and traps \cite{RosHalHug00,SinRetHin0502073}. Here
we describe the fabrication of self-biased Ioffe-Pritchard
microtraps based on FePt film.

The use of permanently magnetized film has many potential
benefits, including very large magnetic field gradients
($\sim$10$^{4}$ T/m). Even for very intricate patterns there is no
ohmic power dissipation, no current noise from power supplies, and
no stray field from lead wires. It is quite conceivable that up to
10$^{5}$ micro traps may be integrated on a square centimeter. An
obvious advantage of current-conducting chips is the possibility
to switch or modulate the currents. As an alternative, we use
external uniform magnetic fields to manipulate the magnetic field
minima.

In this paper we investigate the possibilities of microscopic
patterns of permanent magnetization, in-plane magnetized. In the
second section we briefly discuss the requirements on the magnetic
materials, motivating our choice of FePt alloy. Section 3 is
dedicated to analysis of the patterns that were prepared and
discussions of their application on atom chips. We describe two
types of hard magnetic atom chips. Our trap designs, together with
the actually produced samples, show that in-plane magnetization is
as suitable as out-of-plane magnetization for the purpose of
creating (arrays of) traps. The first design, on a scale of
100~$\mu$m, was cut by spark erosion out of 40 $\mu$m thick FePt
foil. The second design, an array of strips on a scale of
1~$\mu$m, was lithographically patterned in a 250 nm thick film of
FePt on Si.

\section{Material properties}

The material to be used in permanent-magnetic atom chips must meet
a number of requirements. In order to produce a sufficiently large
stray field, a large remanent magnetization M$_{r}$ is required.
Because we use external fields ($\leq$ 10 mT) to move field minima
around we also need a large enough coercivity. Finally, a high
magneto-crystalline an\-iso\-tropy will ensure that the
magnetization is preserved regardless of the shape of the
structure. We found that FePt meets these criteria. FePt has been
studied extensively both in bulk \cite{XiaBruBus04} and thin-film
\cite{WeiSchFah04} form since it combines high magneto-crystalline
anisotropy with high saturation magnetization M$_{s}$
\cite{WelDoe00} and has excellent stability and corrosion
resistance. At high temperature FePt has a disordered
face-centered cubic (fcc) structure which has a very high M$_{s}$
but is magnetically soft. The low-temperature equilibrium
structure on the other hand is face-centered tetragonal (fct or
L1$_{0}$), with a lower M$_{s}$ but very high magneto-crystalline
anisotropy and coercivity. In this phase the Fe and Pt order in an
atomic multilayer structure. It has been shown \cite{LiuLuoLiu98}
that annealing of soft fcc material obtained from a melt produces
nanocrystallites of the fct phase that are exchange coupled to the
soft phase, resulting in a material with the optimum properties
desired here. FePt keeps its magnetic properties for different
thicknesses, both as a bulk and as a film. A unique feature of the
FePt material is that it has a nearly isotropic and large
coercivity, making it suitable for both in-plane and out of plane
field orientation. The former is suitable for the creation of
tight traps, while the latter is suitable for creating waveguides.

We prepared two types of samples. The first consists of a pattern
cut by spark erosion out of bulk Fe$_{0.6}$Pt$_{0.4}$ foil. The
measured ratio M$_{r}$/M$_{s}$ is $\sim$0.8, the saturation
magnetization M$_{s}\approx$ 400 kA/m and the coercivity is about
0.2 T, which is enough to withstand the highest external field
that is applied to manipulate the atoms (10mT). Tests showed that
baking at 170$^{\circ}$C, to obtain UHV conditions, does not
affect the magnetic properties.

The second type of sample consists of an array of strips in 250-nm
thick Fe$_{0.5}$Pt$_{0.5}$ film on Si using lithographic
techniques. For this film thickness the magnetic properties are
essentially similar to the bulk material. The film has been grown
by Molecular Beam Epitaxy (MBE). The in-plane remanence, saturated
magnetization and coercivity of the film are: M$_{r}$/M$_{s}$ =
0.90, M$_{s} \approx$ 600 kA/m and H$_{c}$/$\mu_{0}$ = 0.80 T.
More details on the material optimization and lithographic
processing are given in separate publications
\cite{XinBarGerPP,XinEljGoe04}.

\section{Designing Ioffe Pritchard traps}

\subsection{General remarks}

Magnetic traps of the Ioffe-Pritchard (IP) type
\cite{GotIofTel62,Pri83,BerEreMet87,BagLafPri87} have been used
extensively in the realization of Bose-Einstein condensation of
alkali gases. A crucial property of these traps is that they have
a non-zero magnetic field minimum, in order to prevent Majorana
transitions to untrapped magnetic sublevels. The basic
configuration for an IP trap consists of four long bars with
currents in alternating directions. This generates a cylindrical
quadrupole field, which leads to radial confinement. A set of
axial coils is used to produce a non-zero axial field and to pinch
off the trap along the axis. Atoms in a weak-field seeking spin
state are trapped in the minimum of the magnetic field magnitude.

This basic layout can be implemented with permanent magnets to
create a self biasing structure. Two parallel magnetic strips
produce a cylindrical quadrupole field in a similar way to the
four current carrying bars in a IP trap. The axial field including
the pinch fields are added by placing extra pieces of material in
appropriate places. All dimensions can be scaled down ($<$ 100
nm), resulting in large gradients and curvatures.

The designs described below aim at achieving a trap depth of at
least 0.5 mT, trap frequencies greater than 1 kHz and a non-zero
minimum field so as to avoid spin flipping. The stray field of the
patterns has been calculated using Mathematica \cite{B:Mma50} in
combination with the Radia package \cite{RADIA}.

\subsection{Ioffe Pritchard trap based on FePt foil}

\begin{figure}[t]

\centering{\includegraphics[width=80mm]{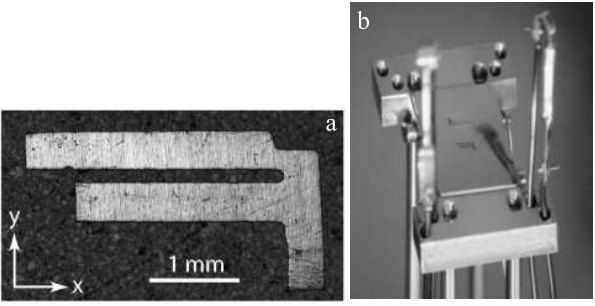}}
\caption{(a): Ioffe-Pritchard trap cut out of a 40 $\mu$m thick
FePt foil, magnetized in-plane ($\parallel y$). (b): Two similar
foil traps glued on an aluminum mirror for mounting inside the
vacuum system. The lower structure corresponds to the left
picture.} \label{fig1}
\end{figure}

A cylindrical quadrupole field can be easily created in the
vicinity of a magnetic pattern, using two identical, uniformly
magnetized strips. The trap is so designed that the two long
strips, in-plane magnetized, produce a cylindrical quadrupole
field with its axis along the $x$ direction, see Fig.~\ref{fig1}.
The bias field needed to lift the magnetic field minimum is given
by the structure itself, in this particular case by the stem of
the "F" and the extension of the upper long strip. These two parts
also determine the trap depth. The structure is thus a self-biased
IP trap. This particular shape is motivated partly by practical
considerations while cutting, and by the desire to have a single
connected piece to simplify mounting on the mirror surface.

\begin{figure}[t]
\centering{\includegraphics[width=60mm]{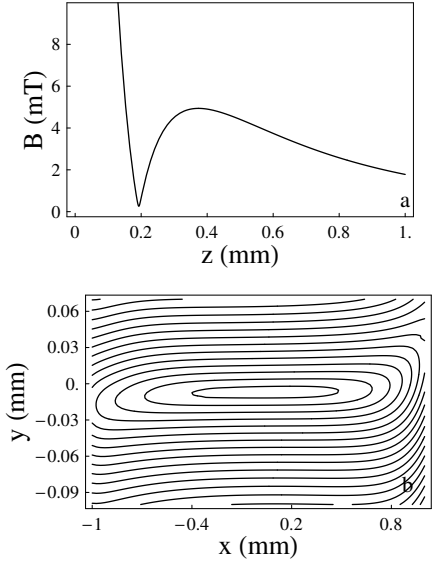}}
\caption{Calculated magnetic field magnitude above the F
structure. (a). Field magnitude versus distance from the chip
surface. The minimum is 0.25 mT. (b) Magnetic field contour lines
of the self-biased IP magnetic trap in the $xy$ plane, in
multiples of 0.5 mT .} \label{fig2}
\end{figure}

We have glued two such magnetic structures with slight\-ly
different dimensions on an aluminum mirror shown in
Fig.~\ref{fig1}(b). The ultra-high vacuum compatible glue should
be used with restraint, since excess glue will cause undesired
scattering of light and hamper absorption imaging of trapped
atoms. The surface of the FePt foil also causes some diffuse light
scattering, which is why it appears somewhat darker than the
mirror in Fig. 1(b). The surface area of the F structures is
sufficiently small that it does not affect the operation of a
mirror magneto-optical trap (MMOT). The resulting magnetic field
pattern produced by one of the structures, Fig.~1(a), is shown in
Fig.~\ref{fig2}. The field strength as a function of the distance
to the surface is shown in Fig.~\ref{fig2}(a). The magnetic field
contour lines in the $xy$ plane are shown in Fig.~\ref{fig2}(b),
marking the B-field strength in multiples of 0.5 mT. A local field
minimum of 0.25 mT is found at a height $z$ = 0.19 mm above the
surface. Both figures 2(a) and 2(b) are cuts through the position
of minimum field. The calculated axial and radial trap
frequencies, for $^{87}$Rb in the F = m$_{f}$ = 2 state, are 51 Hz
and 6.8 kHz, respectively and the trap depth is about 1.1 mT (760
$\mu$K). The second structure, upper in Fig.~\ref{fig1}(b) forms a
trap of 34 Hz in the axial direction and 11 kHz in the radial
direction with a trap depth of 3.3 mT (2.3 mK).

The magnetic foil was prepared from bulk nanocrystalline
Fe$_{0.6}$Pt$_{0.4}$ material. The material was rolled to a 100
$\mu$m foil and mechanically polished to 40 $\mu$m thickness.
Spark erosion using a 50 $\mu$m wire was used to cut the FePt
foil. The size of the gap between the two strips is determined by
the diameter of the cutting wire.

\begin{figure}[t]
\centering{\includegraphics[width=60mm]{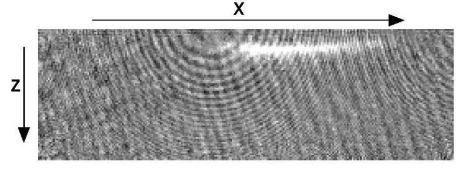}} \caption{An
atomic cloud magnetically trapped in the magnetic field gradient
of  the structure, 400 $\mu$m under the chip surface using an
external field of 6 mT in the x direction. The field of view is
3.3$\times$1.3 mm$^{2}$.} \label{fig3}
\end{figure}

The mirror was mounted inside the vacuum cell with the glued
magnetic structures facing downward. We have trapped 2$
\times10^{6}$ $^{87}$Rb atoms in a mirror magneto optical trap
(MMOT) using the field gradient of the magnetic structures and an
external field of 0.2 mT along $y$. The MMOT forms $\sim$ 2 mm
under the chip. This proves that the sample is still magnetized
after baking of the vacuum setup. Closer to the structure the
magnetic field gradient becomes so strong that the atoms can be
magnetically trapped. We have trapped 5$\times10^{5}$ atoms in
this IP trap using an external field of 6 mT. An absorption image
of the atomic cloud at 400 $\mu$m under the chip is shown in
Fig.~\ref{fig3}. The next step will be to switch off the external
field so that the atoms will be trapped in the self biased IP
trap, requiring detailed optimization of the trajectory into the
bias free trap.

\subsection{Arrays of traps based on micro-fabricated FePt thin
films}

\begin{figure}[t]

\centering{\begin{minipage}{40mm}\includegraphics[width=40mm]{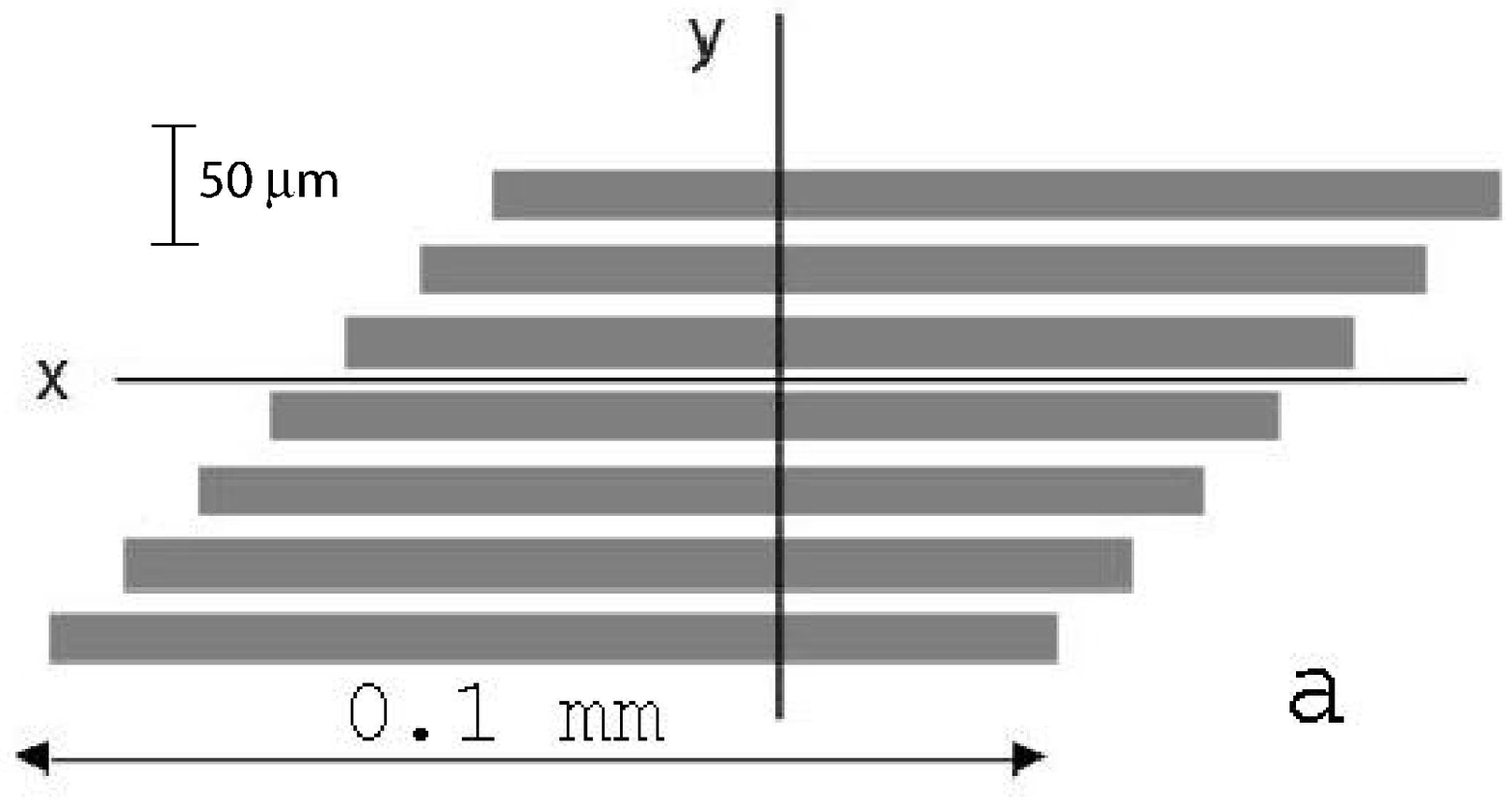}\end{minipage}
\hspace{0cm}
\begin{minipage}{40mm}\includegraphics[width=40mm]{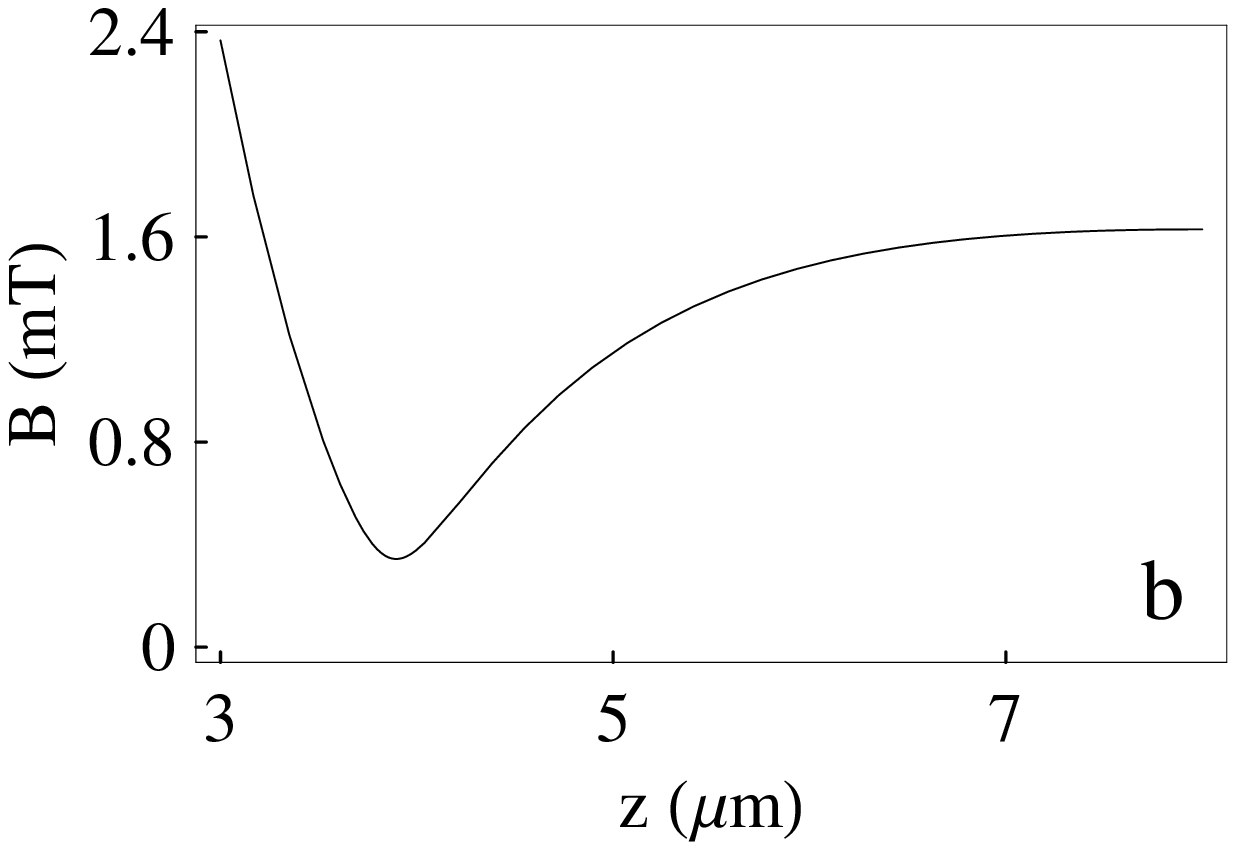}\end{minipage}}
\centering{\begin{minipage}{40mm}\includegraphics[width=40mm]{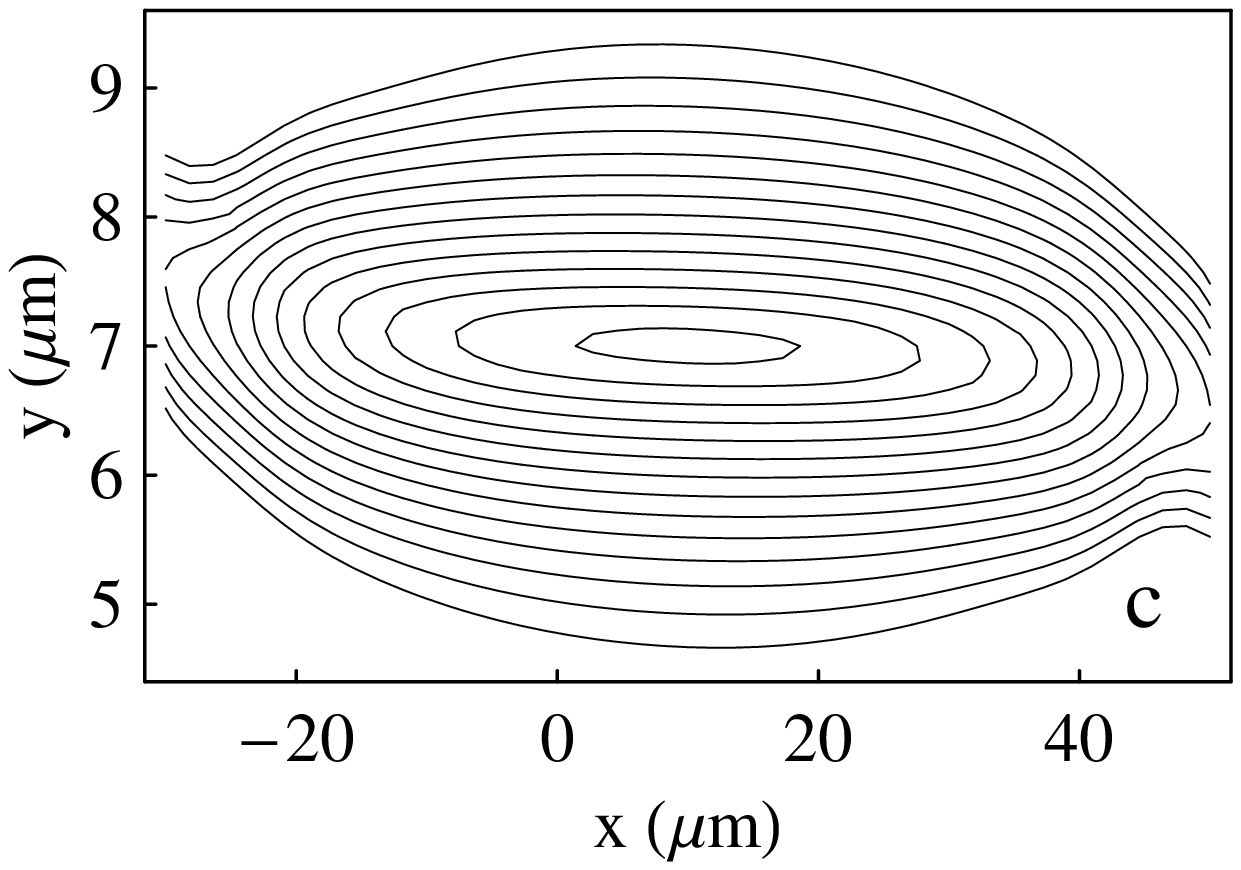}\end{minipage}
\hspace{0cm}
\begin{minipage}{40mm}\includegraphics[width=40mm]{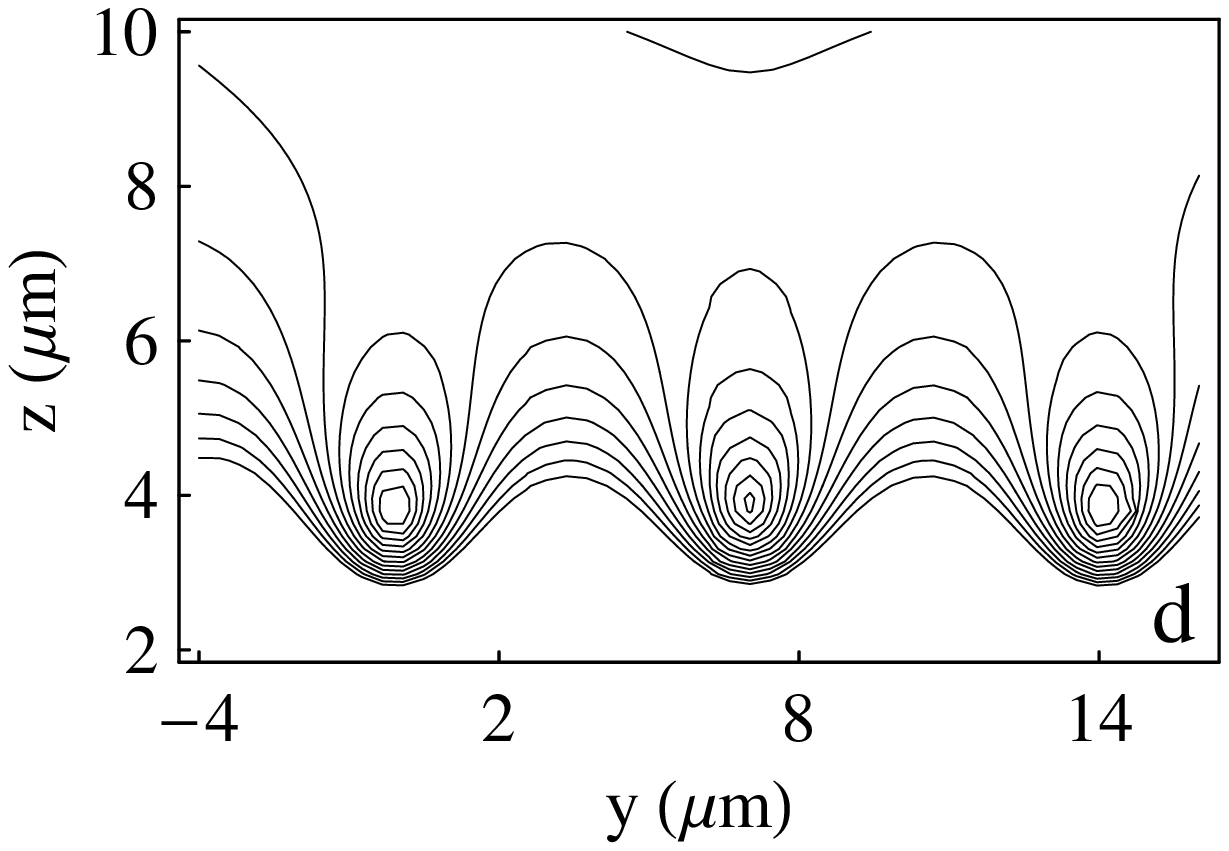}\end{minipage}}
\caption{Calculated array of micro-traps above the surface. (a)
Each strip has a width of 5 $\mu$m and a length of 100 $\mu$m. The
slit is 2 $\mu$m and the longitudinal ($x$) shift is 10 $\mu$m.
(b) Field strength versus $z$ through the center of the trap. (c)
Contours of constant field strength in the $xy$ plane. The
magnetic field contour lines are at 0.3 mT intervals. (d) Contours
of magnetic field magnitude in a $yz$ plane showing an array of
magnetic traps.} \label{fig4}
\end{figure}

In a second approach we designed arrays of magnetic strips
creating arrays of magnetic traps. Such arrays should enable one
to move atoms from one trap to another as in a shift register
practical for quantum information processing. This purpose
requires traps on the micrometer scale, implying films with
(sub-)micrometer thickness. Calculations show that a thickness of
250 nm is satisfactory for making sufficiently tight magnetic
trapping potentials. Each array is characterized by the length and
width of the strips, the width of the slit between strips, and the
longitudinal displacement of neighboring strips as can be seen in
Fig.~\ref{fig4}. By keeping the thickness of the magnetic material
constant and varying the other parameters we can achieve different
trapping potentials at different heights above the surface with
the possibility to reach frequencies in the range of 10 - 100 kHz.
Higher frequencies, in the MHz range could be achieved by further
reduction of the dimensions.

The working principle of this design can be explained as follows:
(n-1) cylindrical quadrupoles are formed between n equally long
strips. As one can see neighboring strips are shifted with respect
to each other by a constant step, like a staircase (see Fig.4(a)).
As a result, for every pair of neighboring strips the upper strip
extends a little to the right of the pair, while the lower strip
extends to the left. These extending pieces of magnetic material
pinch off the ends of each trap and produce the axial field. This
results in a trap with nonzero field minimum being formed between
each pair of neighboring strips, in this case 19 traps using 20
strips. We observe that the trap depth increases with the increase
in the size of these ``end caps''. One example of magnetic field
patterns above the film are presented in Fig.~\ref{fig4}. A plot
of the magnetic field contours reveals a periodic array of
magnetic field minima as shown in Fig 4(d). The trap frequencies
calculated for the central trap are 110 kHz in the radial
direction and 1.5 kHz in the axial direction. The traps are 1.4 mT
(980 $\mu$K) deep. The traps are formed at 4 $\mu$m from the
surface.

\begin{figure}[t]
\centering{\includegraphics[width=60mm]{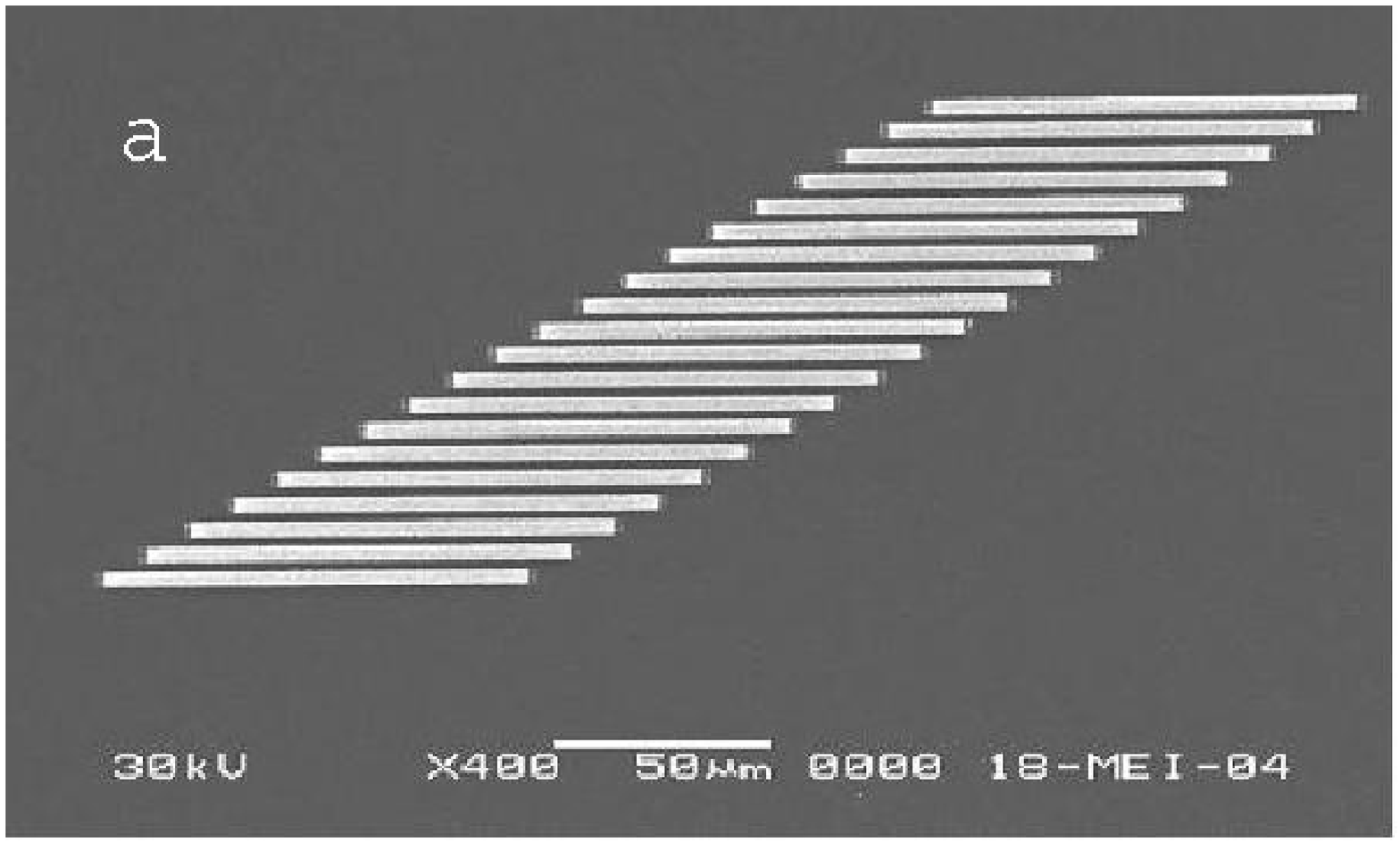} \vspace*{5mm}\\
\includegraphics[width=60mm]{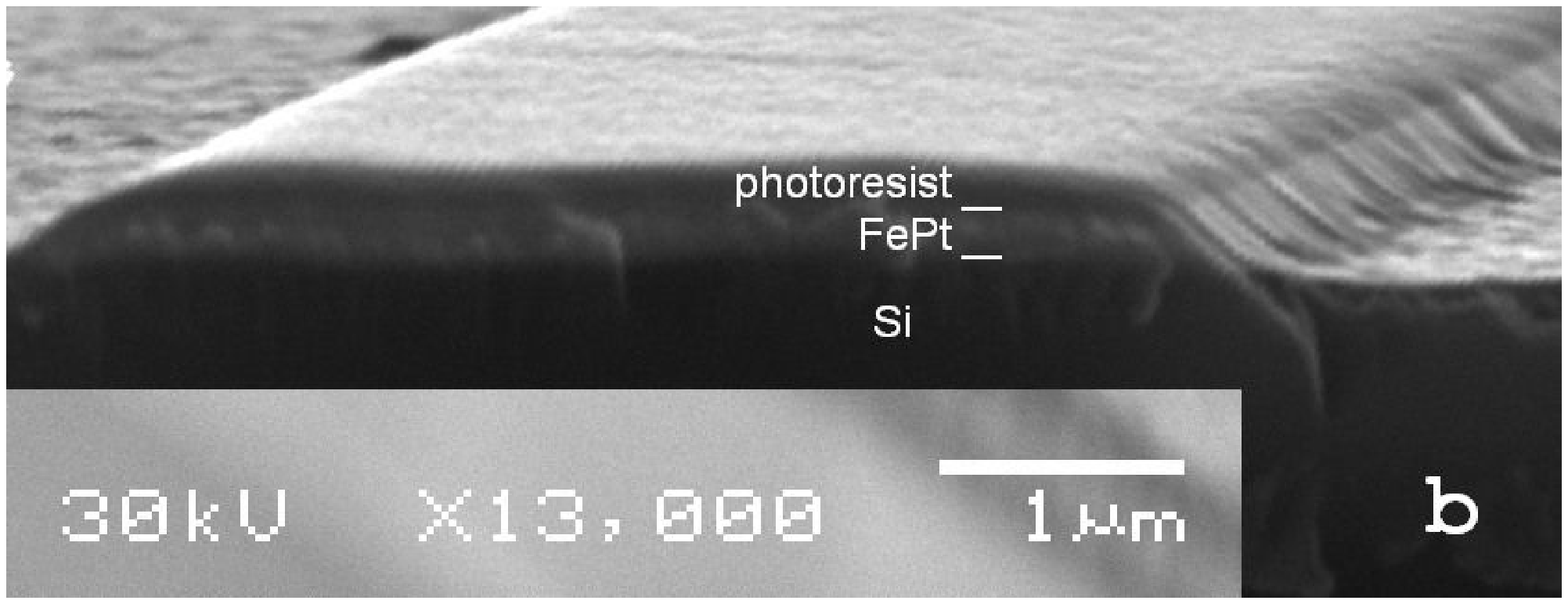}}
\caption{(a)An array of strips of FePt on a silicon substrate and
(b)cross section of one of the FePt strips.} \label{fig5}
\end{figure}

The patterning has been done by e-beam lithography and plasma
etching. Fig.~\ref{fig5} shows SEM images of an array of these
patterns and a cross section of one of them. This shows that the
etching process has yielded an edge at approximately 45$^{\circ}$
angle, with a roughness of approximately 50~nm, which is on the
order of the nanocrystalline grain size. This edge roughness is
slightly larger than the state of the art in current-conducting
chips and can probably still be improved. The edge roughness has
recently been shown to be a critical feature for creating
waveguides in current-conducting chips. An analysis of the
requirements for permanent-magnet chips is in progress. For both
types of chips the requirements for creating traps should be less
stringent, depending on the aspect ratio of the traps.  The final
chip will be coated with a thin Pt layer to obtain a reflecting
surface allowing the formation of the MMOT, and so the photoresist
remaining on top of the FePt pattern will have no effect.

\section{Discussion and conclusions}

We have shown that self-biased micron sized Ioffe-Pritchard traps
and arrays of traps can be prepared with permanent magnetic
structures with in-plane magnetization. We have presented two
types of structures using FePt as magnetic material and two types
of chips have been prepared using foil (40 $\mu$m thickness) and
films (250 nm thickness), with diverse geometries. The bulk
structure and the arrays of strips are suitable for producing
neutral atom traps with high frequencies. The strong magnetization
of FePt in combination with the small structures produce large
magnetic field gradients of order 10$^{4}$ T/m. The FePt film can
be patterned down to nm scales, giving extremely tight traps with
100 kHz, possibly MHz, frequencies for Rb atoms,and making this
method promising for quantum information processing. Previous
theoretical \cite{HenPotWil99} and experimental
\cite{LinTepVul04,JonValHin03,HarMcGCor03} results have shown that
at short distances of several 10 $\mu$m to the surface the
lifetime can be reduced by thermally-induced spin relaxation. The
spin flip lifetime depends on the film thickness, atom-surface
distance and the skin depth \cite{SchRekHin0501149,RekSchHin04}.
For our films of 250 nm thickness, with an estimated skin depth
$\delta$ =100 $\mu$m at 2.1 MHz it should be possible to create
long lived traps at $\gtrsim$ 2 $\mu$m distance from the surface.
The large and near-isotropic coercivity of FePt makes it suitable
for both in-plane and out of plane magnetization. These
microstructures with permanent-magnetic films offer a large
flexibility in design and length scales ranging from mm to (sub-)
$\mu$m feature sizes. A highly promising future approach will be
to develop hybrid chips, combining magnetic structures with other
techniques, including electrostatic potentials, optics, etc.

\begin{acknowledgement}

We thank H. Schlatter, H. Luigjes and C. R\'{e}tif for expert
technical assistance and support. This work was made possible by
the fabrication and characterization facilities of the Amsterdam
$nano$Center. This work is part of the research program of the
Stichting voor Fundamenteel Onderzoek van de Materie (Foundation
for the Fundamental Research on Matter) and was made possible by
financial support from the Nederlandse Organisatie voor
Wetenschappelijk Onderzoek (Netherlands Organization for the
Advancement of Research). This work was supported by the EU under
contract MRTN-CT-2003-505032.

\end{acknowledgement}


\begin{thebibliography}{10}

\bibitem{FolKruCas00}
R. Folman {\it et~al.}, Phys.\ Rev.\ Lett.\ {\bf 84},  4749
(2000).

\bibitem{ReiHanHan99}
J. Reichel, W. H{\"a}nsel, and T. H{\"a}nsch, Phys.\ Rev.\ Lett.\
{\bf 83},
  3398  (1999).

\bibitem{FolKruHen02}
R. Folman {\it et~al.}, Adv.\ At.\ Mol.\ Opt.\ Phys.\ {\bf 48},
263  (2002).

\bibitem{Rei02}
J. Reichel, Appl.\ Phys.\ B {\bf 75},  469  (2002).

\bibitem{DekLeeLor00}
N. Dekker {\it et~al.}, Phys.\ Rev.\ Lett.\ {\bf 84},  1124
(2000).

\bibitem{TreHomRei04}
P. Treutlein {\it et~al.}, Phys.\ Rev.\ Lett.\ {\bf 92},  203005
(2004).

\bibitem{Q:CalHinZol00}
T. Calarco {\it et~al.}, Phys. Rev. A {\bf 61},  022304  (2000).

\bibitem{Q:JanVidCir03}
E. Jan{\'e} {\it et~al.}, Quant.\ Inf.\ \& Comp.\ {\bf 3},  15
(2003).

\bibitem{HanHomHan01}
W. H{\"a}nsel, P. Hommelhoff, T.~W. H{\"a}nsch, and J. Reichel,
Nature {\bf
  413},  498  (2001).

\bibitem{OttForSch01}
H. Ott {\it et~al.}, Phys.\ Rev.\ Lett.\ {\bf 87},  230401
(2001).

\bibitem{SidMcLHan01}
A. Sidorov {\it et~al.}, C.\ R.\ Acad.\ Sci.\ Paris, S{\'e}rie IV
{\bf 2},  565
   (2001).

\bibitem{Dav99}
T.~J. Davis, J.\ Opt.\ B: Quantum Semiclass.\ Opt.\ {\bf 1},
  408  (1999).

\bibitem{SidMcLHan02}
A. Sidorov {\it et~al.}, Acta Physica Polonica B {\bf 33},  2137
(2002).

\bibitem{EriRamHin04}
S. Eriksson {\it et~al.}, Appl.\ Phys.\ B {\bf 79},  811  (2004).

\bibitem{RosHalHug00}
P. Rosenbusch {\it et~al.}, Appl.\ Phys.\ B {\bf 70},  709
(2000).

\bibitem{SinRetHin0502073}
C.~D.~J. Sinclair {\it et~al.}, arXiv:physics/0502073  .

\bibitem{XiaBruBus04}
Q. Xiao {\it et~al.}, J.\ All.\ Comp.\ {\bf 364},  315  (2004).

\bibitem{WeiSchFah04}
M. Weisheit, L. Schultz, and S. F{\"a}hler, J.\ Appl.\ Phys.\ {\bf
95},  7489
  (2004).

\bibitem{WelDoe00}
D. Weller and M. Doerner, Ann.\ Rev.\ Mat.\ Sc.\ {\bf 30},  611
(2000).

\bibitem{LiuLuoLiu98}
J. Liu, C. Luo, Y. Liu, and D. Sellmyer, Appl.\ Phys.\ Lett.\ {\bf
72},  483
  (1998).

\bibitem{XinBarGerPP}
Y.~T. Xing {\it et~al.}, unpublished  .

\bibitem{XinEljGoe04}
Y.~T. Xing {\it et~al.}, Phys.\ Stat.\ Sol.\ (c) {\bf 12},  3702
(2004).

\bibitem{GotIofTel62}
Y. Gott, M. Ioffe, and V. Tel'kovskii, Nucl.\ Fusion, 1962 Suppl.\
{\bf Pt.\
  3},  1045  (1962).

\bibitem{Pri83}
D.~E. Pritchard, Phys.\ Rev.\ Lett.\ {\bf 51},  1336  (1983).

\bibitem{BerEreMet87}
T. Bergeman, G. Erez, and H.~J. Metcalf, Phys.\ Rev.~A {\bf 35},
1535  (1987).

\bibitem{BagLafPri87}
V. Bagnato {\it et~al.}, Phys.\ Rev.\ Lett.\ {\bf 58},  2194
(1987).

\bibitem{B:Mma50}
{Wolfram Research, Inc.}, {\em Mathematica, Version 5.0}
(Champaign, IL, 2003).

\bibitem{RADIA}
O. Chubar, P. Elleaume, and J. Chavanne, J.\ Synchroton Rad.\ {\bf
5},  481
  (1998), the Radia code is freely available for download from
  http://www.esrf.fr/machine/support/ids/Public/Codes/\\software.html.

\bibitem{HenPotWil99}
C. Henkel, S. P{\"o}tting, and M. Wilkens, Appl.\ Phys.\ B {\bf
69},  379
  (1999).

\bibitem{LinTepVul04}
Y. Lin, I. Teper, C. Chin, and V. Vuleti{\'c}, Phys.\ Rev.\ Lett.\
{\bf 92},
  050404  (2004).

\bibitem{JonValHin03}
M.~P.~A. Jones {\it et~al.}, Phys.\ Rev.\ Lett.\ {\bf 91},  080401
(2003).

\bibitem{HarMcGCor03}
D. Harber, J. McGuirk, J. Obrecht, and E. Cornell, J.\ Low Temp.\
Phys.\ {\bf
  133},  229  (2003).

\bibitem{SchRekHin0501149}
S. Scheel, P. Rekdal, P. Knight, and E. Hinds,
arXiv:quant-ph/0501149  .

\bibitem{RekSchHin04}
P.~K. Rekdal, S. Scheel, P.~L. Knight, and E.~A. Hinds, Phys.\
Rev.~A {\bf 70},
   013811  (2004).

\end{thebibliography}

\end{document}